\documentclass[conference]{IEEEtran}
\IEEEoverridecommandlockouts
\usepackage{cite}
\usepackage{amsmath,amssymb,amsfonts}
\usepackage{algorithmic}
\usepackage{graphicx}
\usepackage{graphbox}
\usepackage{subfig}
\usepackage{textcomp}
\usepackage{xcolor}
\usepackage{url}

\def\BibTeX{{\rm B\kern-.05em{\sc i\kern-.025em b}\kern-.08em
    T\kern-.1667em\lower.7ex\hbox{E}\kern-.125emX}}
\begin{document}

\title{A Ferroelectric Tunnel Junction-based Integrate-and-Fire Neuron
\thanks{This work was supported in part by the European Union's Horizon 2020 Research and Innovation Programme under Grant Agreement No 871737 and in part by the European Research Council (ERC) through the European's Union Horizon Europe Research and Innovation Programme under Grant Agreement No 101042585. Views and opinions expressed are however those of the authors only and do not necessarily reflect those of the European Union or the European Research Council. Neither the European Union nor the granting authority can be held responsible for them.\\
© IEEE 2022. Personal use is permitted. Permission must be obtained from IEEE for all other uses, in any form of media, including reprinting/republishing the material, for advertising or promotional material, creating new collective works, for resale or redistribution to servers or lists, or reuse of any copyrighted component of this work in other works.}
}


\author{\IEEEauthorblockN{P. Gibertini\IEEEauthorrefmark{2}\IEEEauthorrefmark{1},
L. Fehlings\IEEEauthorrefmark{2}\IEEEauthorrefmark{1},
S. Lancaster\IEEEauthorrefmark{2}, 
Q. T. Duong\IEEEauthorrefmark{2},
T. Mikolajick\IEEEauthorrefmark{2}\IEEEauthorrefmark{3},
C. Dubourdieu\IEEEauthorrefmark{4}\IEEEauthorrefmark{5},
S. Slesazeck\IEEEauthorrefmark{2},\\
E. Covi\IEEEauthorrefmark{2}, and
V. Deshpande\IEEEauthorrefmark{4}}
\IEEEauthorblockA{\IEEEauthorrefmark{2}NaMLab gGmbH, Dresden, Germany \IEEEauthorrefmark{3}Technical University of Dresden, Dresden, Germany}
\IEEEauthorblockA{\IEEEauthorrefmark{4}Helmholtz Zentrum Berlin, Berlin, Germany \IEEEauthorrefmark{5}Free University Berlin, Berlin, Germany}
\IEEEauthorblockA{\IEEEauthorrefmark{1}These Authors contributed equally to this work}
\IEEEauthorblockA{Corresponding Authors: erika.covi@namlab.com, veeresh.deshpande@helmholtz-berlin.de}}

\maketitle

\begin{abstract}
Event-based neuromorphic systems provide a low-power solution by using artificial neurons and synapses to process data asynchronously in the form of spikes. Ferroelectric Tunnel Junctions (FTJs) are ultra low-power memory devices and are well-suited to be integrated in these systems. Here, we present a hybrid FTJ-CMOS Integrate-and-Fire neuron which constitutes a fundamental building block for new-generation neuromorphic networks for edge computing. We demonstrate electrically tunable neural dynamics achievable by tuning the switching of the FTJ device.
\end{abstract}

\begin{IEEEkeywords}
Integrate-and-Fire Neuron, FTJ, HZO, neuromorphic computing, edge computing.
\end{IEEEkeywords}

\section{Introduction}
Event-based neuromorphic systems where data is processed asynchronously as spikes are an attractive solution for low power edge computing. The basic computing blocks in such systems are artificial neurons and synaptic circuits. While there has been significant effort to develop CMOS neurons and synapse circuits~\cite{covi2021FrontNeurosci}, the recent developments in emerging devices provides an opportunity for integration with CMOS technology to design neuron and synaptic circuits that can be power and area efficient compared to their CMOS only counterparts~\cite{fleps22}. Among the various emerging devices, ferroelectric tunnel junctions (FTJ) feature ultra-low switching energy~\cite{max2019JEDS}. Hafnium Zirconium Oxide (Hf${_{0.5}}$Zr${_{0.5}}$O${_2}$; HZO) based FTJ devices can be integrated on the CMOS back-end-of-line enabling hybrid FTJ-CMOS circuits~\cite{deshpande2021}. Despite growing interest in synaptic applications of HZO FTJs~\cite{max2020hafnia}, neuron circuits have not been much explored. The HZO FTJ device features thresholding behavior in resistance change which is useful for neuron circuits. In this context, we present a hybrid FTJ-CMOS Integrate-and-Fire neuron circuit. The circuit utilizes an HZO-based bilayer FTJ device that realizes the pulse integration. The circuit can become a building block for low-power neuromorphic systems.   

\begin{figure}[!t]
    \centering
    \subfloat[]{
        \includegraphics[width=0.9in]{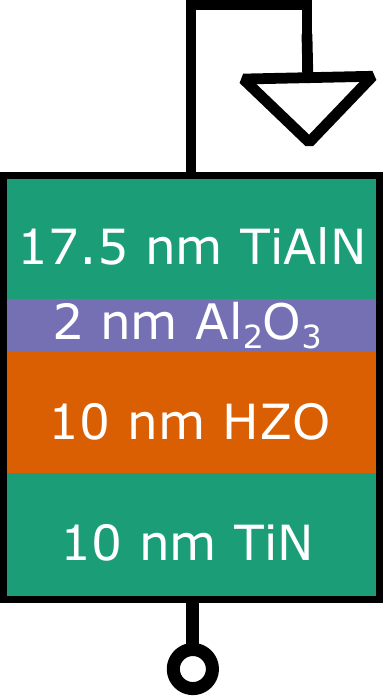}
        \label{fig:schematic}
    }
    \subfloat[]{
    \includegraphics[width=2.2in]{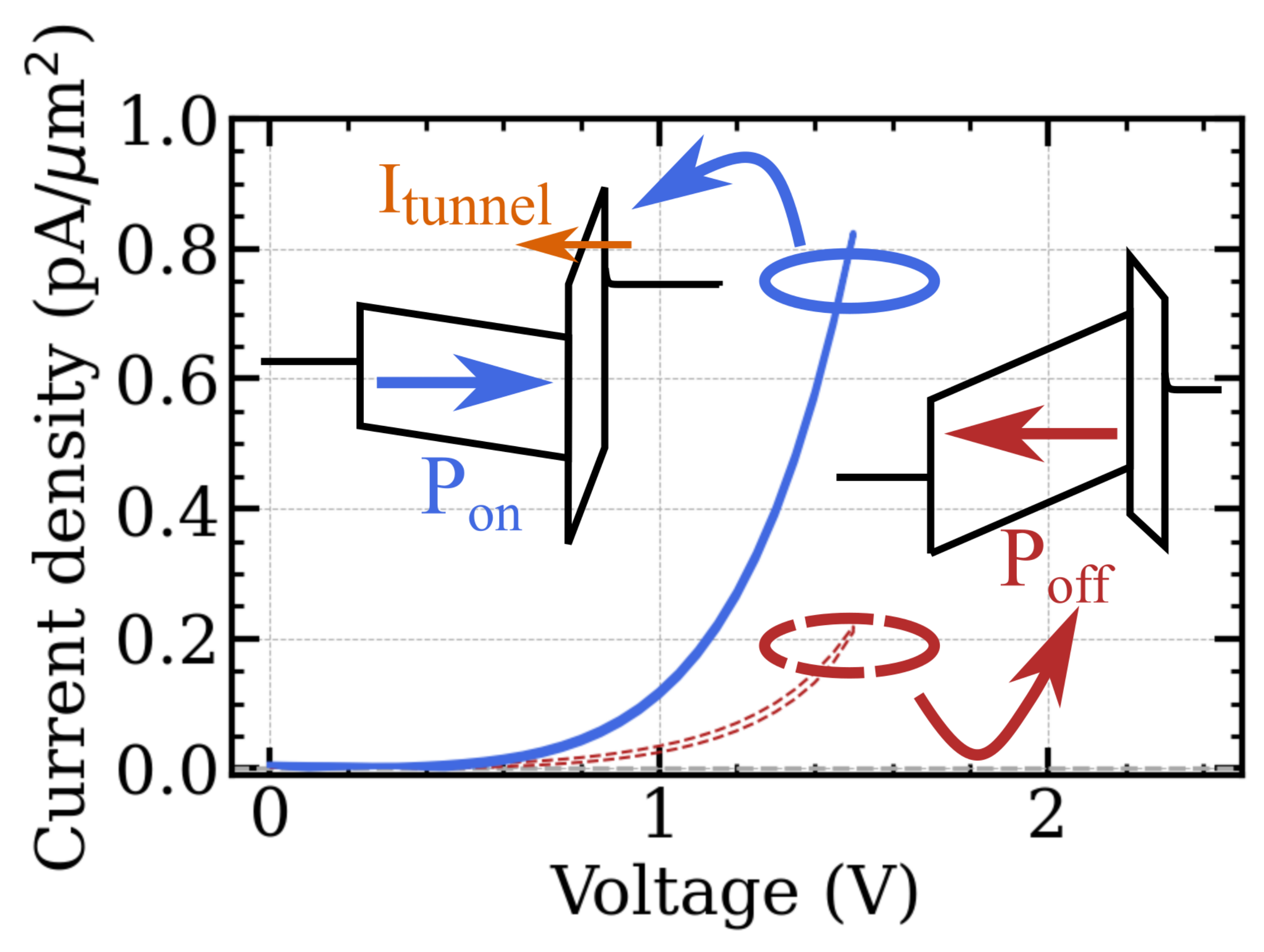}
    \label{fig:onoff}
    }
    \caption{\protect\subref{fig:schematic} Schematic of the FTJ stack. \protect\subref{fig:onoff} DC-IV measurements of the FTJ in ON and OFF state, measured up to 1.5\,V. Schematics of the corresponding band structures are shown in the inset.}
    \label{fig:ftj_schematic}
\end{figure}

\section{Gradual switching properties of FTJ devices}
\label{sec:device}
\begin{figure}[!t]
    \centering
    \includegraphics[width=2.5in]{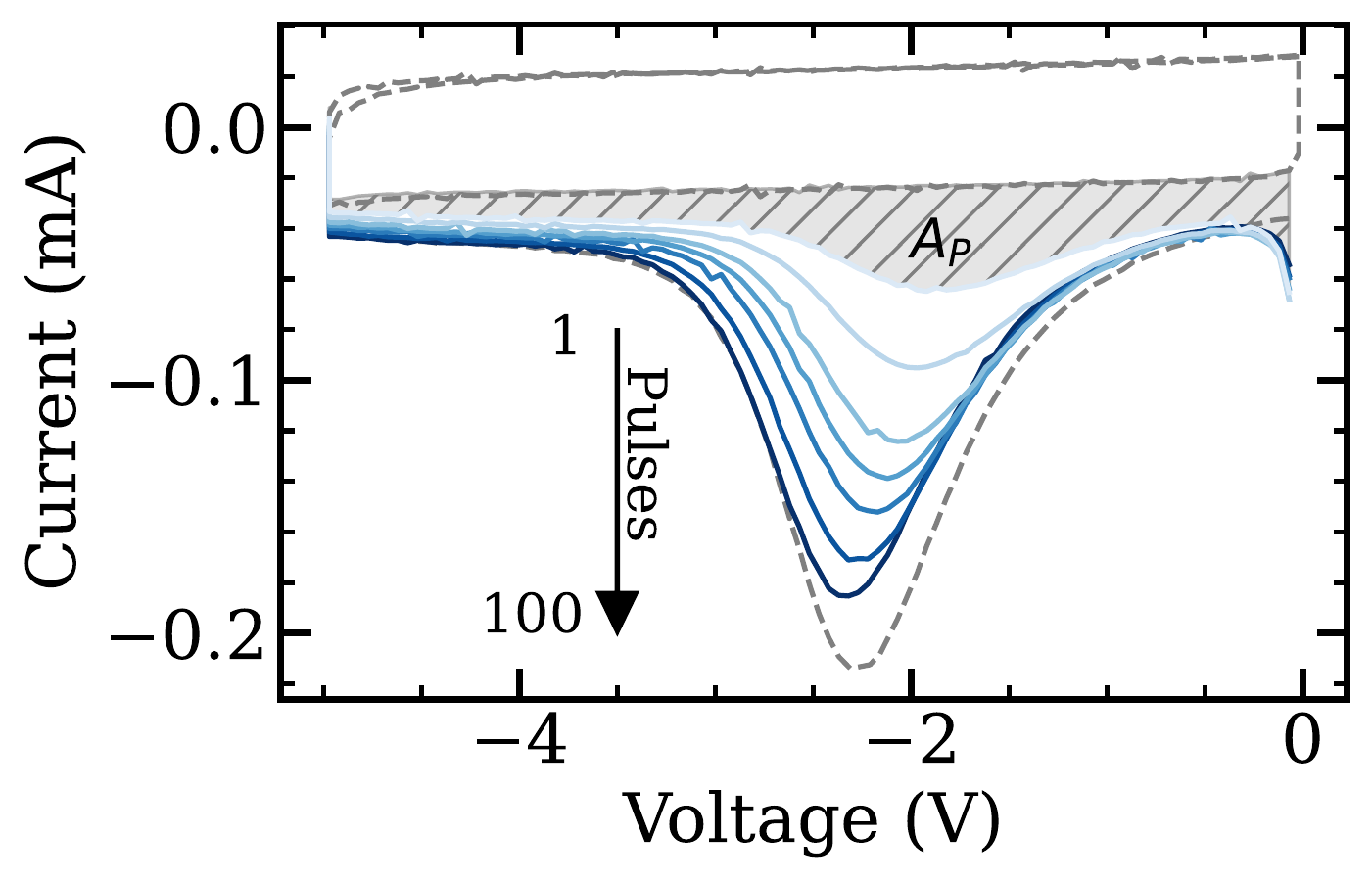}
    \caption{Measured FTJ reset currents after an exponentially increasing amount of pulses with amplitude 4\,V and length 10\,\textmu s. The shaded area $A_P$ is integrated to obtain the polarization in Fig. \ref{fig:switched_polarization}. The dashed line represents the complete switching process with a single -5\,V, 500\,\textmu s triangular voltage pulse.}
    \label{fig:switching_currents}
\end{figure}

\begin{figure}[!t]
    \centering
    \includegraphics[width=2.5in]{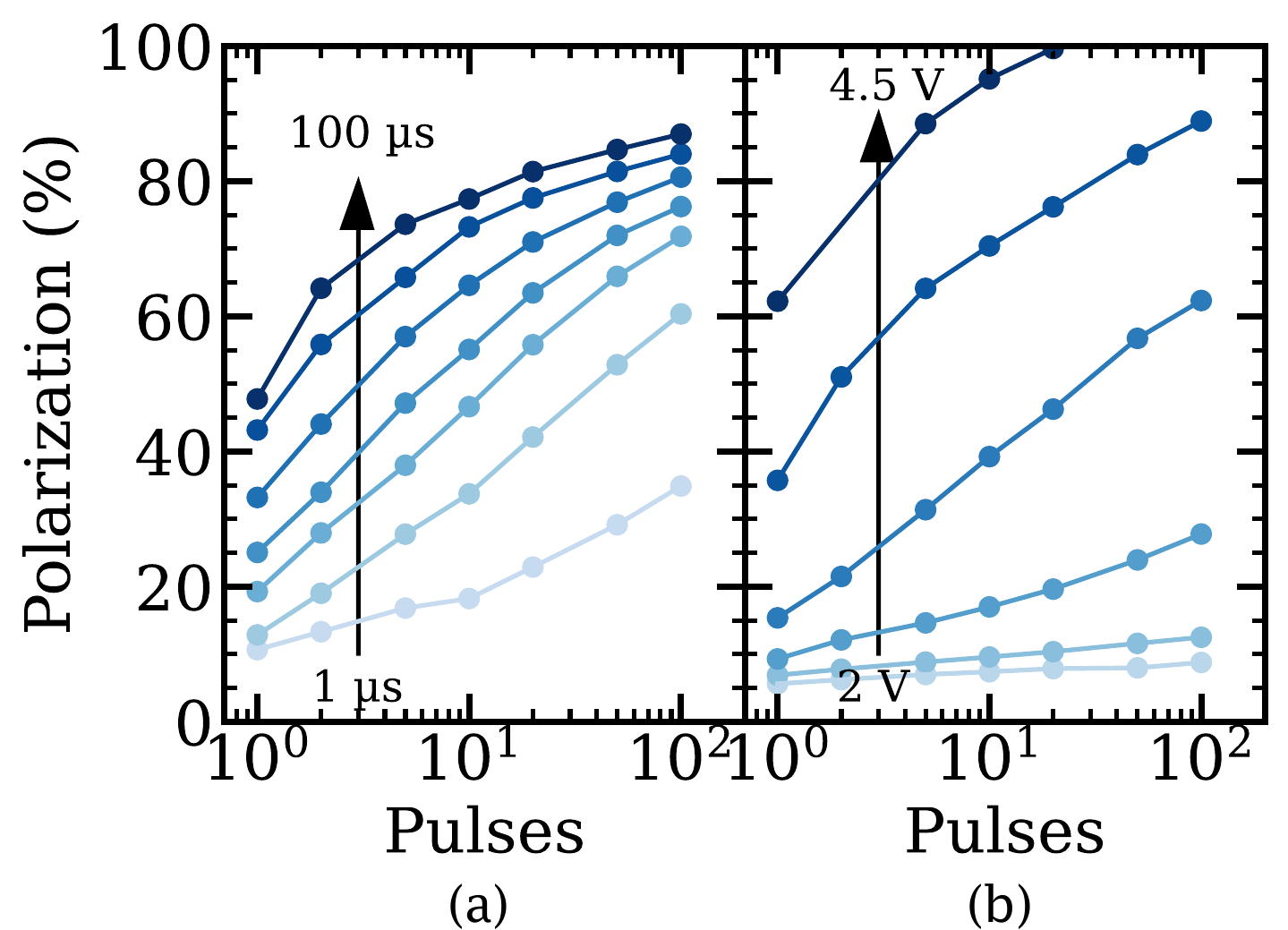}
    \caption{Measured switched polarization as a function of the number of applied pulses, for (a) exponentially increasing pulse widths at 3.5\,V and (b) different voltages in steps of 0.5\,V at a pulse width of 10\,\textmu s. The polarization is normalized to a full switching measurement.}
    \label{fig:switched_polarization}
\end{figure}

Fig. \ref{fig:schematic} shows the HZO based bilayer FTJ device considered in this work. Details of the FTJ device fabrication can be found in \cite{lancaster2022essderc}. The operation principle of the devices is shown through band diagrams in the insets of Fig.~\ref{fig:onoff}. When the ferroelectric polarization (blue/red arrows) points towards the Al$_2$O$_3$ layer, an electric field is induced which results in the ON state defined by a high tunneling current. When the polarization points away, the band bending results in a larger tunneling barrier width and an OFF state with a low tunneling current.
The polarization is switched ON (set) by applying 5\,V, and switched OFF (reset) by applying -5\,V. The polarization can be partially switched by using lower voltage amplitudes or shorter pulse times. The currents of these intermediate states are directly proportional to the amount of switched polarization~\cite{lancaster2022multi}. 
In order to study gradual switching under identical set or reset pulses (representing spikes) for the neuron integration function, the switching under multiple square pulses with fixed amplitude and width was characterized. Variable numbers of voltage pulses were applied to the device and the switched polarization was measured by switching the device back with a single, triangular -5\,V, 500\,\textmu s pulse~\cite{lancaster2022multi}. Fig. \ref{fig:switching_currents} shows the back-switching current after gradual switching with pulses of 4\,V amplitude and 10\,\textmu s width. Based on the non-switching current measured on a second back-switching pulse, the switching current and accordingly the switched polarization can be measured, which is illustrated by the shaded area in Fig. \ref{fig:switching_currents}. Complete switching was measured under a triangular voltage pulse with amplitude 5\,V and width 500\,\textmu s. An evaluation of the switched polarization for different pulse widths and an amplitude of 3.5\,V (Fig. \ref{fig:switched_polarization}a), as well as for different voltages with a pulse width of 10\,\textmu s (Fig. \ref{fig:switched_polarization}b) supports the viability of the gradual switching approach employed in the neuron circuit.


\section{FTJ Compact Model}
\label{sec:model}

\begin{figure}[!t]
    \centering
    \subfloat[]{
    \includegraphics[align=c,width=2.3in]{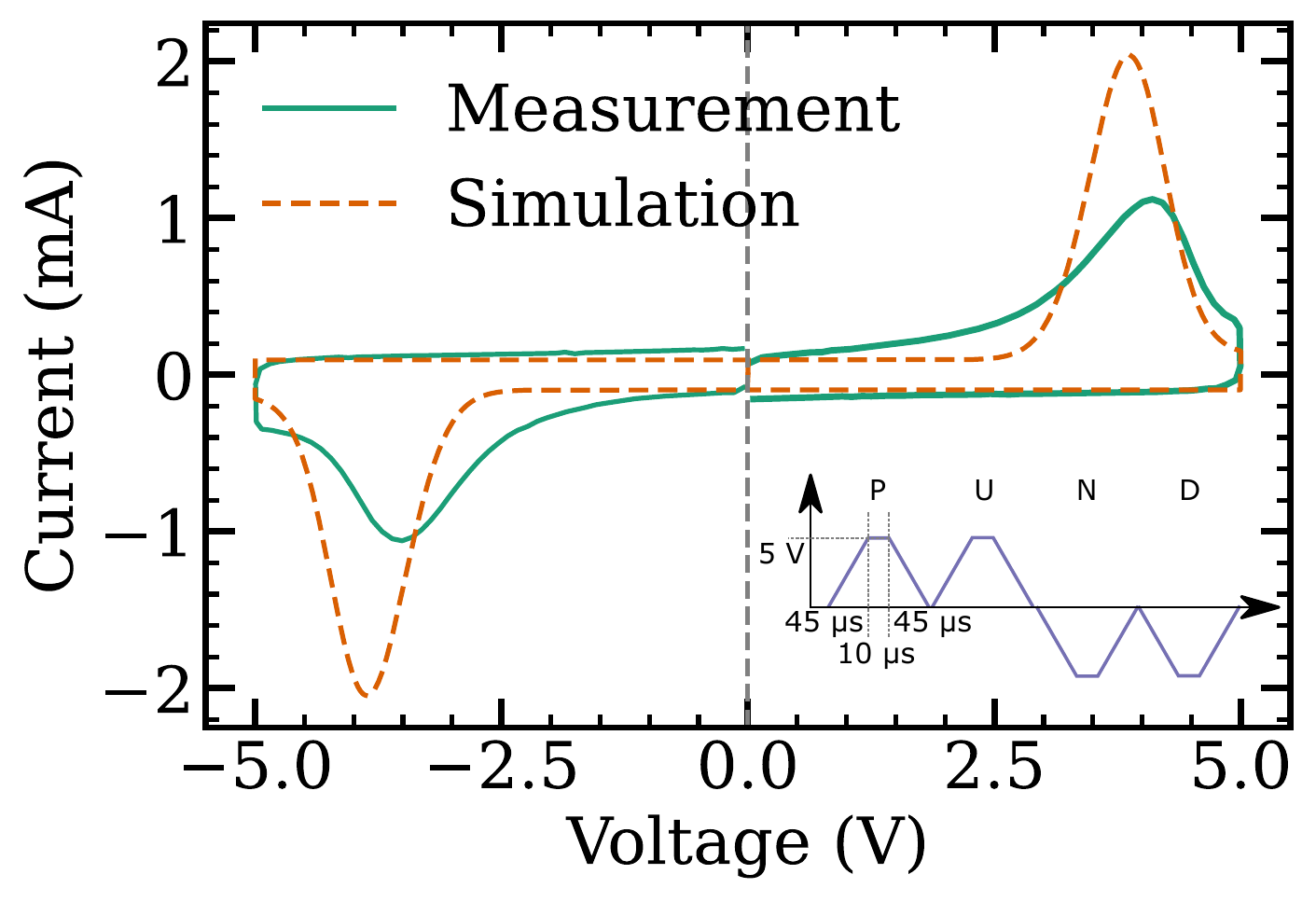}
    \label{fig:pund}
    }
    \subfloat[]{
    \includegraphics[align=c,width=1in]{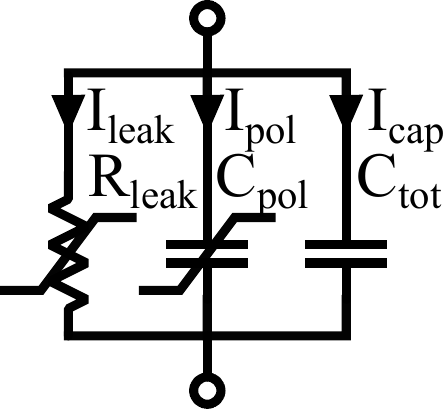}
    \label{fig:model_ftj}
    }
    \caption{(a) Switching I-V characteristic obtained with 100\,\textmu s wide trapezoidal excitation (shown inset) and the equivalent simulated I-V obtained using the compact model. (b) Equivalent circuit.}
    
\end{figure}

For the circuit simulations, a compact model based on a HZO ferroelectric capacitor was used. Here we provide a summary of the salient aspects of the model. For a more detailed description, please see~\cite{BeFerro_D2.4}.
We do not model a two-layer structure, rather we take the Al$\mathrm{_2}$O$\mathrm{_3}$ layer into account by adapting the value of the effective coercive field $E_C$ to represent the capacitive divider we have in our FTJ. The HZO layer consists of a mix of ferroelectric, antiferroelectric, and paraelectric phases due to the fabrication process. Therefore, the total capacitance of the FTJ is modeled as the parallel of a ferroelectric capacitor and a dielectric one (Fig.~\ref{fig:model_ftj}).

The FTJ model is based on the Preisach model of ferroelectric hysteresis and has been calibrated to fit the characteristics of the device discussed in Section~\ref{sec:device}. The Preisach model describes the evolution of the polarization $P$ as a function of applied electric field $E$ 
~\cite{preisach1935springer}. We calculate the asymptotic polarization for a certain voltage following Miller et al. ~\cite{miller1990JAP} as:

\begin{equation}
    P(E,k,P_{off}) = k\cdot P_{sat}\cdot tanh\left(\frac{E_{eff}\pm E_C}{2\delta}\right)+P_{off}
\label{eq:P}
\end{equation}

where $k$ and $P_{off}$ are a scaling factor and the offset polarization, respectively, used to account for unsaturated loops~\cite{jiang1997VLSI}, $P_{sat}$ is the saturation polarization, $E_{eff}$ the field applied to the ferroelectric layer, and $\delta$ is:

\begin{equation}
    \delta = E_C \left(ln\frac{1+P_r/P_{sat}}{1-P_r/P_{sat}}\right)^{-1}
\end{equation}

where $P_r$ is the remanent polarization.
The switching kinetics is added to the model by using the static polarization $P$ as in (\ref{eq:P}) and adding a time dependence to find the dynamic polarization $P_{dyn}$:

\begin{equation}
    P_{dyn} = \frac{P\,\Delta t + P_{old} \cdot \tau_{pE}}{\tau_{pE} + \Delta t}
\end{equation}

$\Delta t$ is the actual time-step of the simulation, $P_{old}$ is the polarization calculated in the previous time-step, and $\tau_{pE}$ is

\begin{equation}
    \tau_{pE} = \tau_p\,10^{\frac{E_C - abs\left(E_{eff}\right)}{\alpha_E\left(\frac{E_C}{10}+abs\left(E_{eff}\right)\right)}}
\end{equation}

where $\tau_p$ and $\alpha_E$ are fitting parameters.

Each change in the polarization results in a polarization current, $I_{pol}$, modeled through the ferroelectric capacitor (Fig. \ref{fig:model_ftj}):
\begin{equation}
    I_{pol} = \frac{d P_{dyn}}{d t} \cdot A_{tot}
\end{equation}

The actual state of the FTJ is given by a current, $I_{leak}$, flowing through a non-linear resistor placed in parallel to the capacitors. The current through this resistor depends on the polarization state of the ferroelectric layer (Fig. \ref{fig:model_ftj}):

\begin{equation}
    I_{leak} = R_{A0}\,A_{tot}\,\left(e^{\frac{V}{V_{PE}}}-1\right)
\end{equation}

where $R_{A0}$ is a fitting constant dependent on the ferroelectric material stack, $A_{tot}$ is the area of the FTJ, $V$ is the applied voltage, and $V_{PE}$ is a polarization-dependent variable:

\begin{equation}
    V_{PE} = V_{P0} - \Delta V_P \cdot \frac{P_{dyn}}{P_{sat}}
\end{equation}

with $V_{P0}$ and $\Delta V_P$ fitting constants.

Fig.~\ref{fig:pund} compares a typical measured and simulated I-V characteristic of our FTJ devices measured with 100\,$\mu$s-wide trapezoidal positive-up-negative-down pulses. The simulation was obtained using the values from Table~\ref{tab:param}. The model agrees well with the capacitive behavior and the coercive voltage $V_c$ of the device ($V_c$ is higher than in Fig.~\ref{fig:switching_currents} due to the measurement frequency). The evolution of the polarization, given as the integral of the I-V curves, is roughly the same, but the simulated switching current is higher and narrower than the measured one (about 2.1\,mA peak current obtained in the simulation versus 1.1\,mA measured in the device). Notably, the simulation yields a Gaussian current peak which is not observed experimentally. This is due to the model assumption of a statistically independent switching of domains.

\begin{table}[]
    \centering
    \caption{Values of the parameters used to calibrate the compact model.}
    \begin{tabular}{|c|c||c|c|}
        \hline
        \textbf{Parameter} & \textbf{Value} &  \textbf{Parameter} & \textbf{Value} \\
        \hline
        \hline
        $E_C$ & 3.3\,$MV/cm$ & $\tau_{p}$ & 10\,$\mu s$ \\
        \hline
        $k$ & 0.5 & $\alpha_E$ & 0.25 \\
        \hline
        $P_{sat}$ & 20.0000\,$\mu C/cm^2$ & $A_{tot}$ & 3.14$\cdot$10$\mathrm{^{-4}}$\,$cm^2$ \\
        \hline
        $P_r$ & 19.9997\,$\mu C/cm^2$ & $R_{A0}$ & 110\,$\mu A$ \\
        \hline
        $V_{P0}$ & 0.36\,$V$ & $\Delta V_P$ & 0.06\,$V$ \\
        \hline
    \end{tabular}
    \label{tab:param}
\end{table}

\section{FTJ-based integrate-and-fire neuron}

\begin{figure}[!t]
    \centering
    \includegraphics[width=2.5in]{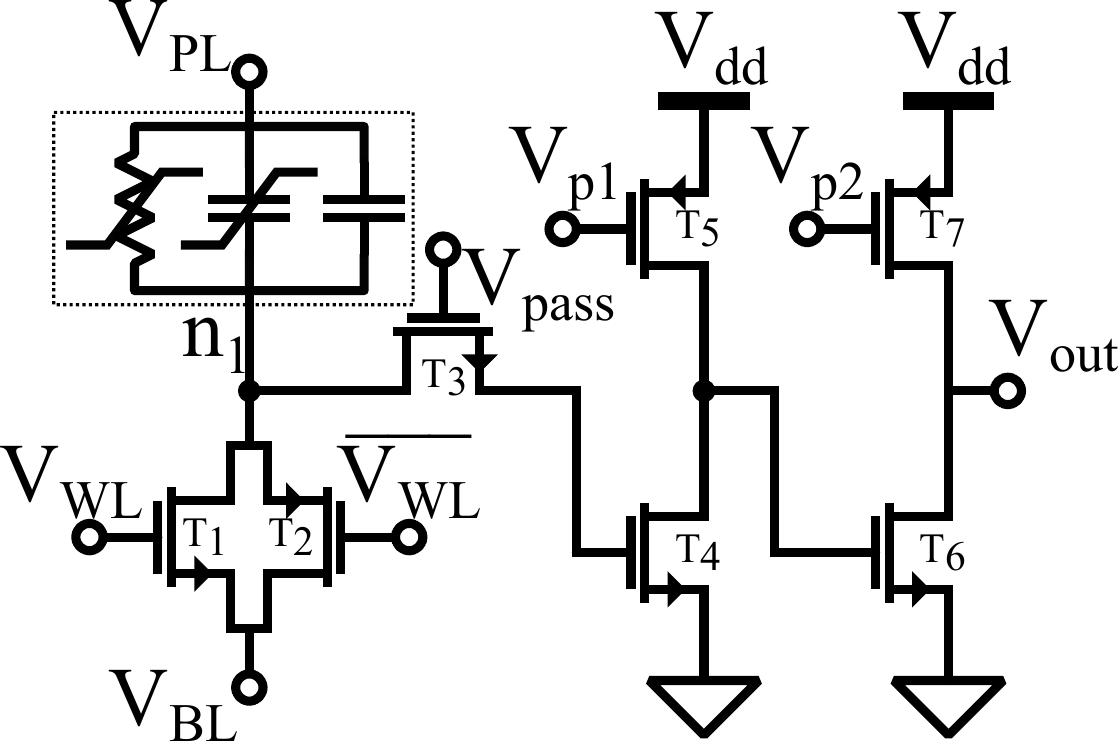}
    \caption{Integrate-and-Fire neuron schematic. Transistors $T_1$ to $T_4$ implement a variation of the 2T1C cell \cite{slesazeck2019uniting}. Transistors $T_4$ to $T_7$ implement two inverters with electrically tunable switching threshold.}
    \label{fig:neuron_schematic}
\end{figure}

\begin{figure*}[!t]
    \centering
    \subfloat[]{
        \includegraphics[height=2.15in]{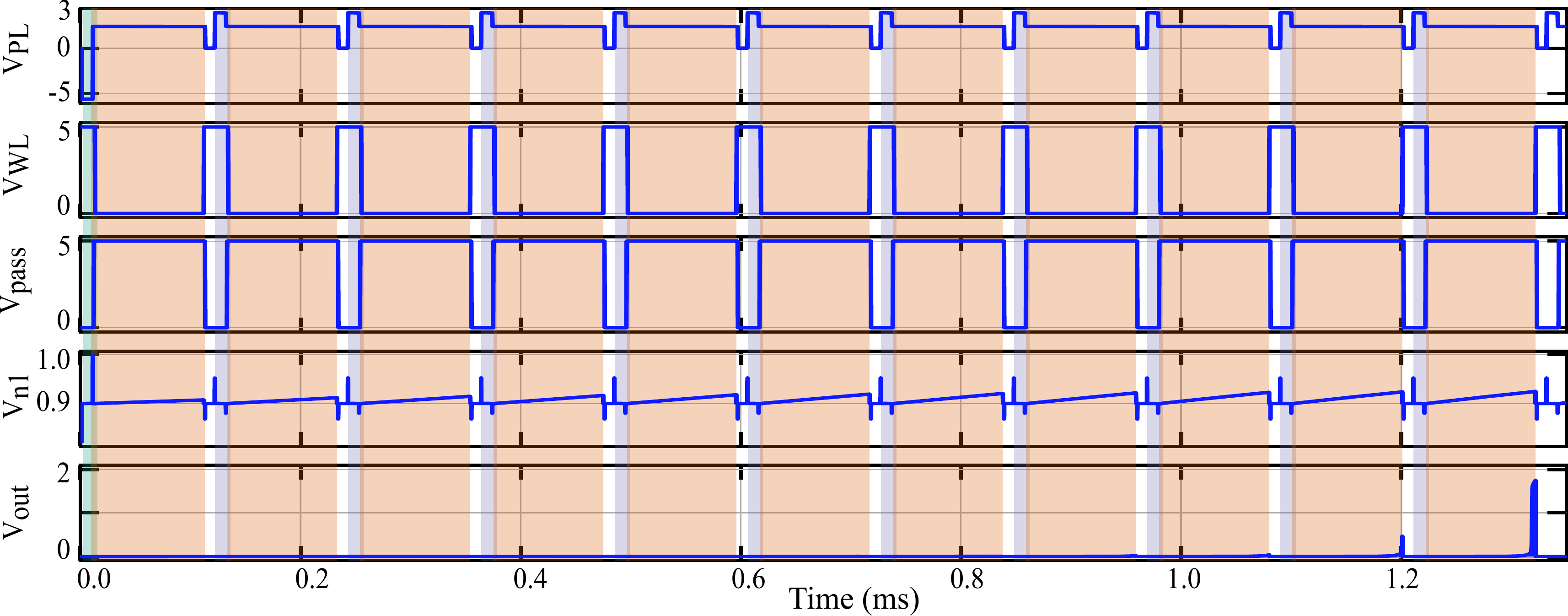}
        \label{fig:serie_pp}
    }
    \subfloat[]{
    \includegraphics[height=2.15in]{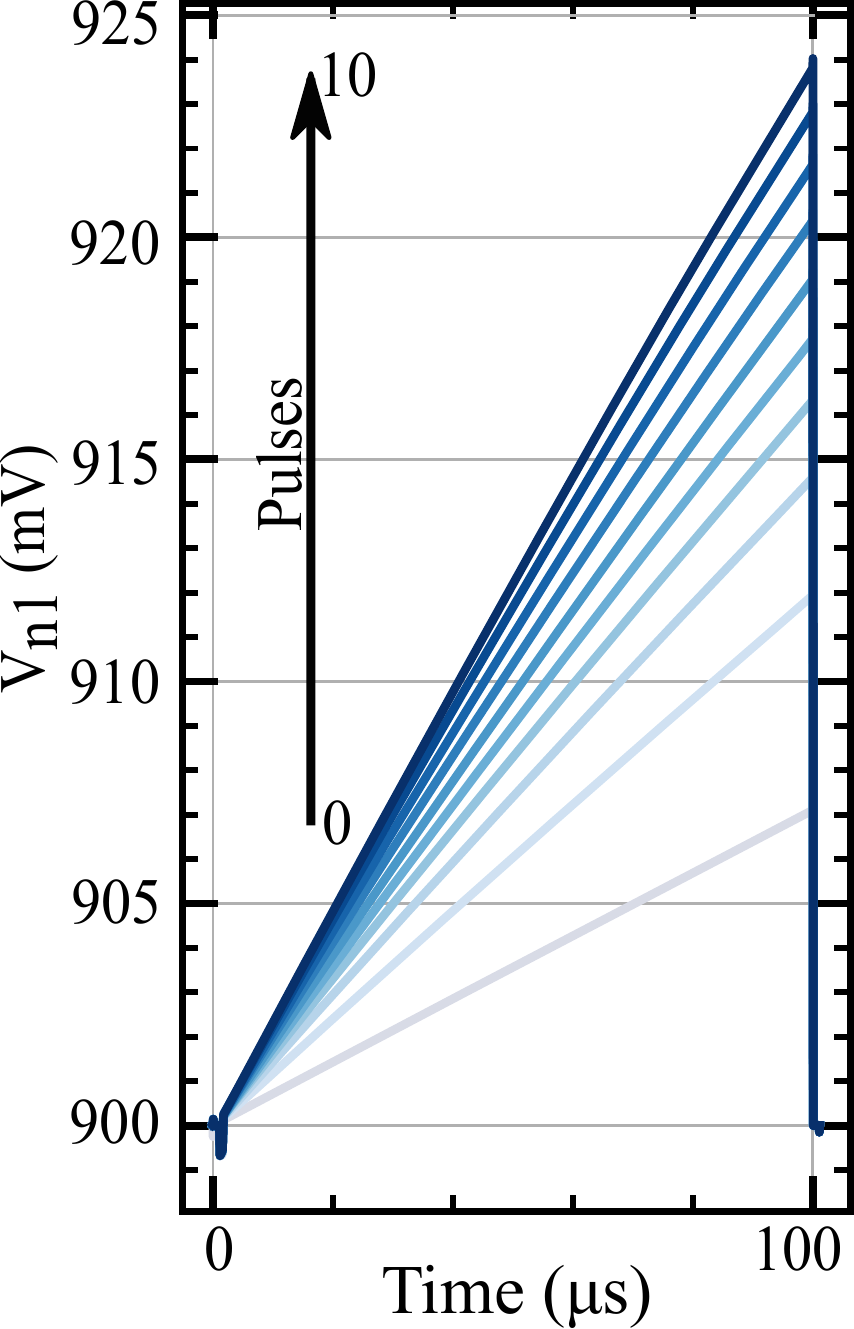}
    \label{fig:n1_zoom}
    }
    \caption{(a) Simulation of the neuron behavior from the initial condition (reset) to the first spike. The green shade highlights the reset operation, the purple the set operation, and the orange the integrating phase. The pre-charge phase occurs where the different colors (green and orange, purple and orange) are overlapping. All the y axes are given in V. (b) Evolution of node $n_1$ during the reading phases shown in (a).}
    \label{fig:timing_diagram}
\end{figure*}

The very low leakage current in FTJ devices (see Fig. \ref{fig:onoff}) allows very energy efficient operations, but poses a challenge in reading the memory state. The proposed circuit, shown in Fig.~\ref{fig:neuron_schematic}, adopts a variation of the 2 transistor-1 capacitor (2T1C) readout scheme ($T_1$ to $T_4$ in Fig.~\ref{fig:neuron_schematic}) which uses a read transistor $T_4$ to amplify the signal from the FTJ \cite{slesazeck2019uniting}. The circuit was designed in a standard 180\,nm CMOS technology. Transistor $T_3$ is added to prevent any disturbance to reach the gate of $T_4$ during the writing phase. Two inverters with tunable switching threshold ($T_4$ to $T_7$) are used in order to obtain sharp pulses starting from a slow changing input signal~\cite{FGneuron}. The PMOS gate voltages ($V_{p1}$ and $V_{p2}$) are used to tune the switching threshold of the inverters.
During the writing operation (green and purple shades in Fig.~\ref{fig:serie_pp}), transistors $T_1,~ T_2$ are turned ON through $V_{WL}$, while the pass transistor $T_3$ is OFF. Voltage pulses, negative to reset and positive to set, are applied at the plate line ($PL$) while the bit line ($BL$) is kept at a constant voltage.
The reading operation (orange shade in Fig.~\ref{fig:serie_pp}) is divided in two different phases, pre-charge and integration. During the pre-charge phase (2\,$\mu$s, evidenced by the overlapping of the orange shade with the green / purple one in Fig.~\ref{fig:serie_pp}), transistors $T_1,~ T_2$, and $T_3$ are turned ON, thus charging node $n_1$ to $V_{BL}$. $V_{BL}$ is chosen near the switching voltage of the first inverter, in the high-gain region. Then the integrating phase begins. $T_1,~ T_2$ are turned OFF so that $BL$ is disconnected, leaving $n_1$ floating, and a read voltage ($V_{read}$) is applied to $PL$. Under these conditions, the FTJ discharges, i.e., the voltage at $n_1$ ($V_{n1}$) increases, because of the FTJ leakage current. Indeed, the device can be modeled as an RC circuit where the capacitance is constant and the resistance is proportional to the polarization (see Sect.~\ref{sec:model}). The final $V_{n1}$ after a read phase of 100\,\textmu s depends on the resistance of the FTJ. If it is higher than the switching voltage of the first inverter, $V_{out}$ rises and the neuron fires, as shown in Fig.~\ref{fig:serie_pp}. Fig.~\ref{fig:n1_zoom} shows $V_{n1}$ during each read phase shown in Fig.~\ref{fig:serie_pp}. The slope of $n_1$ depends on the state of the FTJ and the difference between the final voltage at the end of the reading phase after the first and the last pulse is of almost 20\,mV.

It is worth noting that the signals which provide the full functionality of the neuron (e.g., bias and pulse generators) have to be generated outside the neuron itself, although on the same chip. However, the additional circuit can be shared among several neurons when using an asynchronous, event-based approach. Each single neuron can process event occurring at a minimum time distance of 150\,\textmu s. The limit on the bandwidth imposed by the necessary integration time of the single neuron is overcome in scaled systems with a large number of neurons \cite{moradi2017scalable}.

The choice of $V_{read}$ is critical, as there is only a small window where a readable leakage current is obtained (see Fig.~\ref{fig:onoff}) without modifying the polarization. Therefore, a voltage of 1.5\,V was chosen.
Other important parameters for the performance of the neuron are the $BL$ voltage $V_{BL}$ and the integration, i.e., read, time. Decreasing the integration time and/or reducing $V_{BL}$ slows down the neuron response, i.e., increases the number of pulses needed before the firing event.

\begin{figure}[!t]
    \centering
    \includegraphics[width=2in]{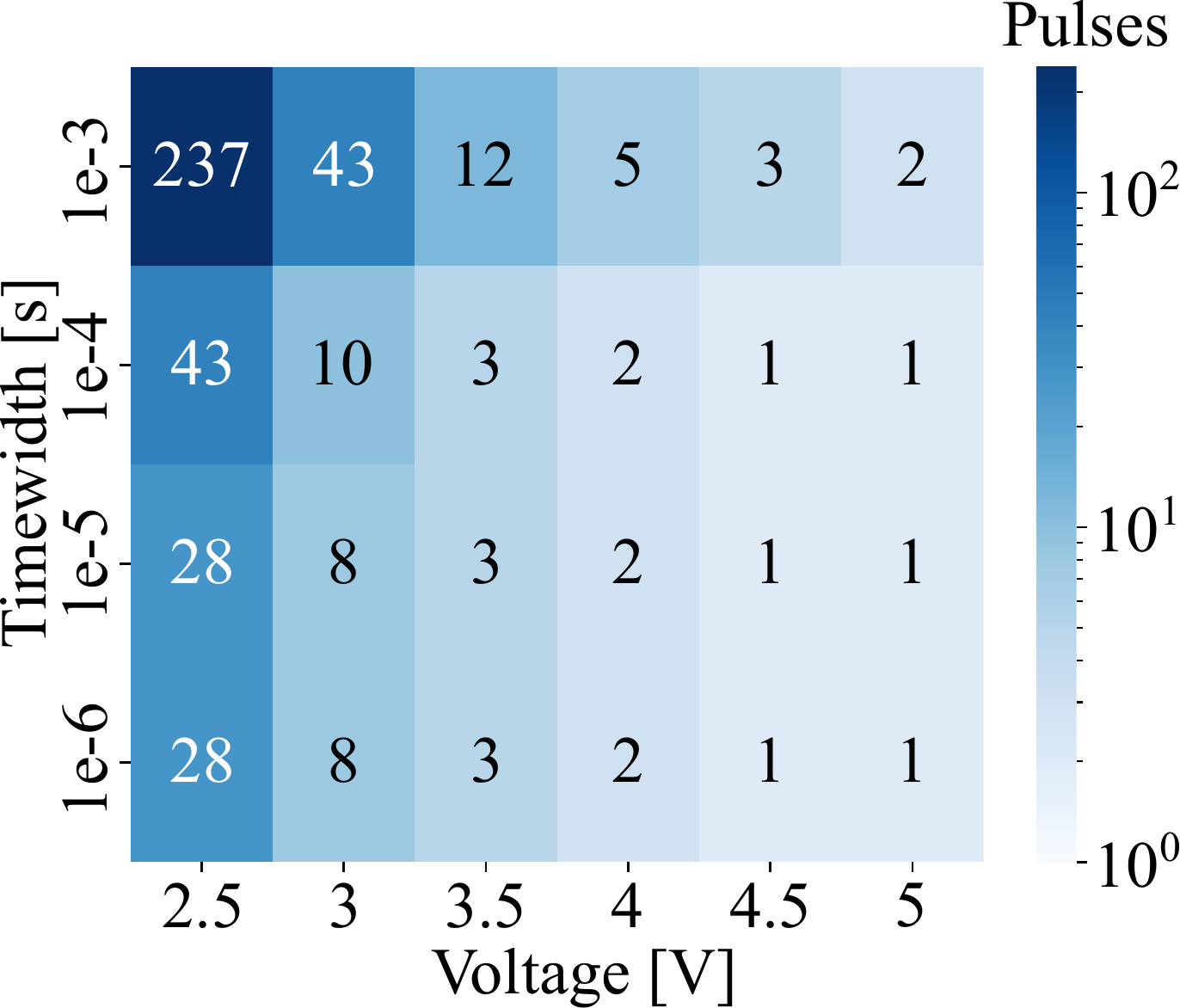}
    \caption{Simulated number of pulses before a firing event as a function of the voltage amplitude and of the time width of the set pulse. The number of pulses is written inside the colored square.}
    \label{fig:npuls_vs_tw_and_vw}
\end{figure}

The reset can be performed by applying a strong negative pulse (-5\,V for 10\,\textmu s) or by applying 5\,V at the $BL$ while keeping the $PL$ at 0\,V. The behavior of the neuron, i.e., the number of pulses needed to fire, can be modulated by changing the parameters of the set pulse. In accordance to the experimental data (Fig.~\ref{fig:switched_polarization}), we can observe in Fig.~\ref{fig:npuls_vs_tw_and_vw} that a change in the set pulse amplitude or time width results in a different number of pulses needed before the firing event. It is observed that higher voltages / longer times lead to firing with a lower number of pulses. From the perspective of circuit design, it would be more convenient to exploit the time- rather than voltage-dependence of the FTJ. However, the FTJ shows a better modulation with voltage amplitude than time width. A change in voltage of 2.5\,V results in a gradual reduction of pulses before fire, from 43 to 1 for 10\,\textmu s pulses. When fixing the voltage at 3\,V and modulating the pulse width, a sharp reduction is instead observed between 10$^{-6}$ and 10$^{-5}$\,s, from 43 to 10 pulses before fire. This hardly changes for pulse widths over the next two orders of magnitude. One way to improve flexibility in the neural behavior and cope with process variability, both at circuit and at device level, whilst still keeping the design compact can be by changing the values of $V_{p1}$ and $V_{p2}$ to change the switching thresholds of the inverters. More specifically, a higher voltage at the gate of $T_5,~ T_7$ lowers the switching threshold of the inverters. 

\section{Conclusions}
In this work, a novel hybrid FTJ-CMOS neuron circuit design and simulations are presented. The integration function in the neuron can be achieved by accumulative polarization switching behavior, which has been demonstrated experimentally in FTJ devices under application of identical pulses. The read current through the FTJ is amplified by the CMOS circuits and leads to the firing event. We simulated the pulse number needed for firing and its dependence on voltage pulse amplitude and width. The possibility of electrically tuning the neural dynamics by controlling the switching of the FTJ makes the proposed neuron suitable as a fundamental building block in new-generation neuromorphic computing systems. 

\bibliographystyle{ieeetr}
\bibliography{refs}

\end{document}